\documentclass[]{iopart}

\usepackage{graphicx}
\usepackage{floatflt}
\usepackage{tabularx}

\def\gtorder{\mathrel{\raise.3ex\hbox{$>$}\mkern-14mu
 \lower0.6ex\hbox{$\sim$}}}
\def\ltorder{\mathrel{\raise.3ex\hbox{$<$}\mkern-14mu
 \lower0.6ex\hbox{$\sim$}}}

\begin{document}

\title{Coulomb corrections in the extraction of the proton radius}

\author{John Arrington}
\address{Physics Division, Argonne National Laboratory, Argonne, IL 60439, USA}

\date{\today}

\begin{abstract}

Multi-photon exchange contributions are important in extracting the proton
charge radius from elastic electron--proton scattering.  So far, only
diagrams associated with the exchange of a second photon have been evaluated.
At the low $Q^2$ values relevant to the radius extraction, and especially 
the very low $Q^2$ region to be probed by proposed measurements, higher order
contributions may become important.  We evaluate these corrections in the
Effective Momentum Approximation, which includes the Coulomb interaction to
all orders, and find small corrections with a strong $Q^2$ dependence at low
$Q^2$ and large scattering angles.  This suggests that the higher order terms
may be important in the evaluation of the proton magnetic radius.

\end{abstract}


\maketitle


The proton RMS charge radius, $R_p$, has become a topic of great interest
following recent Lamb shift measurements in muonic hydrogen~\cite{pohl10,
antognini13} which are extremely sensitive to $R_p$.  These measurements yield
$R_p$=0.8409(4)~fm, significantly smaller than recent extractions based on
electron--proton interactions~\cite{sick03, mohr08, bernauer10, zhan11,
bernauer13}. The extractions
include both electron scattering and electron--proton interactions in the
hydrogen atom, which yield a combined result of $R_p$=0.8772(46)~\cite{zhan11,
ron11}, or $R_p$=0.8775(51) according to the CODATA10 evaluation~\cite{mohr12}.
Significant efforts have gone into examining the current extractions and
examining possible physics which could explain the discrepancy~\cite{pohl13},
but there is not yet a clear explanation.

In the atomic physics measurements, atomic transitions in hydrogen (or muonic
hydrogen) are measured with high precision.  Because the non-relativistic
wavefunction for the electrons in S states have some overlap with the finite
charge distribution of the proton, the Coulomb potential is modified at very
short distances. This yields a very small correction to these energy levels
and thus in transition such as the Lamb shift (2S-2P transition), which has
been measured to to better than a part in 10$^{14}$.  Extraction of the radius
from this measurement involves many corrections to the non-relativistic
calculation, including uncertainty associated with the details of the spatial
distribution of charge in the proton.  Including these uncertainties, the
radius extracted from such measurements is 0.8758(77)~fm (Adjustment 6 of
Ref.~\cite{mohr12}).

In muonic hydrogen, the muon has significantly more overlap with the charge
distribution of the proton because it is approximately 200 times more massive
than the electron.  Thus, the impact of the finite size of the proton yields a
much greater correction to the energy levels, so a significantly less precise
measurement of the Lamb shift in muonic hydrogen can determine the proton
radius to much greater precision: $R_p = 0.84087(39)$~fm~\cite{antognini13}. 
Again, there are significant corrections~\cite{eides01, karshenboim05,
antognini13b} to the simple non-relativistic calculations, which have been
examined in great detail in light of the discrepancy~\cite{pohl13}.

In elastic electron-proton scattering, the finite size of the proton yields
a deviation in the cross section from scattering off of a point-like charge.
These are encoded in the charge and magnetic form factors, $G_E(Q^2)$ and
$G_M(Q^2)$, where $-Q^2$ is the square of the four-momentum transfer between
the electron and proton, and thus represents the momentum scale at which the
proton structure is being probed.  In the non-relativistic picture, the
rest-frame RMS charge radius is directly related to the slope of the charge
form factor at $Q^2$=0. However, relativistic boost corrections need to be
applied, breaking the equivalence of these quantities.  Because these boost
corrections depend on the complicated sub-structure of the protons, the
convention is to define the radius through the relation
\begin{equation}
R_p^2 = -6 \frac{dG_E(Q^2)}{dQ^2}
\end{equation}
in the limit $Q^2 \to 0$.  Note that this is the equivalent to the definition
of $R_p$ used in the extraction from the Lamb shift in hydrogen and atomic
hydrogen.

Radiative corrections play an important role in the electron scattering
measurements. While the largest corrections are well understood, the
diagrams which depend on the proton structure, e.g. the two-photon exchange
(TPE) diagrams, cannot be calculated exactly due to the hadronic structure
uncertainty.  In the past, these corrections were calculated in the 2nd Born
approximation, assuming the exchange of a second soft photon with an unexcited
intermediate state.  Initial calculations were performed in the limit $Q^2 \to
0$~\cite{mckinley48}, and later at finite $Q^2$, which require a
parameterization of the proton charge and magnetic form factors, $G_E$ and
$G_M$.  The inclusion of these corrections was found to be important in the
extraction of the proton radius~\cite{rosenfelder00, sick03}.

More recently, calculations going beyond the 2nd Born approximation have been
performed~\cite{maximon00, blunden05a, afanasev05a, borisyuk06a, borisyuk08,
kivel09, borisyuk09}, motivated by the discrepancy between Rosenbluth and
polarization measurements at high $Q^2$~\cite{arrington03a, guichon03,
arrington07a, perdrisat07, arrington11a}. See recent reviews~\cite{carlson07,
arrington11b} for further details on the different theoretical approaches.
There have also been several phenomenological extractions of TPE
contributions~\cite{arrington04d, borisyuk07b, chen07, belushkin08, graczyk11,
qattan11a, qattan11b} and attempts to experimentally constrain
TPE~\cite{qattan05, tvaskis06, VEPP-3, arrington09b, borisyuk10,
guttmann11, meziane11, moteabbed13}, but essentially all of these works focus
on $Q^2 > 2$--3~GeV$^2$, where there is a clear discrepancy between the
Rosenbluth and polarization extractions of the form factors.

\begin{figure}[!htbp]
\begin{center}
\includegraphics*[width=3.1in]{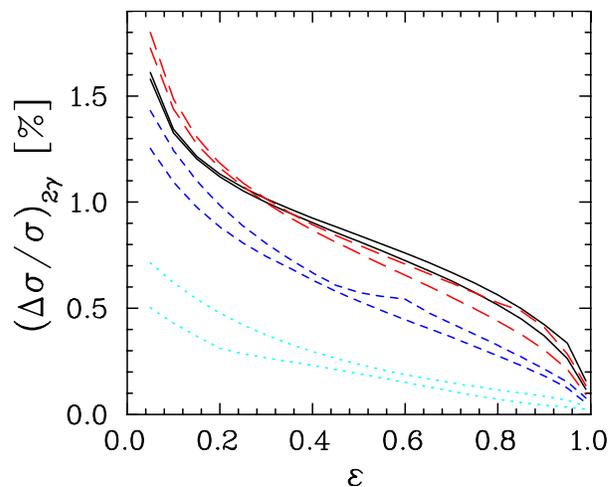}
\end{center}
\caption{Range of results for several TPE calculations at low
$Q^2$~\cite{blunden05a, kondratyuk05, borisyuk06a, borisyuk07a, borisyuk08,
borisyuk12, borisyuk13}.
Solid, long-dashed, short-dashed, and dotted lines show the range for
$Q^2$=0.01, 0.03, 0.1, and 0.2~GeV$^2$, respectively.
Note that the low $Q^2$ expansion~\cite{borisyuk07a} is only valid up to
$Q^2$=0.1~GeV$^2$, and so is excluded from the
$Q^2=0.2$~GeV$^2$ range.}
\label{fig:tpe_compare}
\end{figure}

\begin{figure}[!htbp]
\begin{center}
\includegraphics*[width=3.1in]{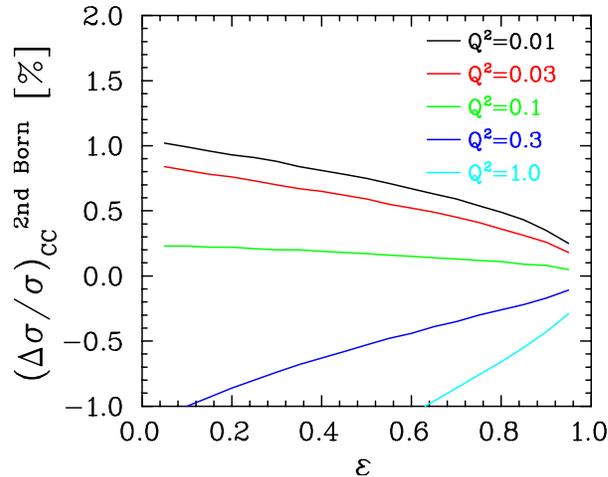}
\end{center}
\caption{Coulomb corrections calculated in the 2nd Born
approximation for e--p elastic scattering for $Q^2$ values from 0.01~GeV$^2$
(top curve) to 1~GeV$^2$ (bottom curve).}
\label{fig:cc_2ndborn}
\end{figure}

All of these calculations include only the effect of a second virtual photon
exchange.  Because the corrections are at the few percent level and this
is an expansion in the fine structure constant, $\alpha \approx 1/137$, it
is expected that two-photon terms should be sufficient.  However, while higher
order terms in the expansion are expected to be small, the approximations made
in evaluating these contributions can yield a large range of results.
The full TPE calculations include both the exchange of hard and soft photons,
while evaluations in the 2nd Born approximation include only the exchange 
of a second soft photon.  One can see that this has a significant impact
even at low $Q^2$, and especially at low $\varepsilon$, by comparing the
results in Figures~\ref{fig:tpe_compare} and~\ref{fig:cc_2ndborn}.  In some
cases, e.g. Ref.~\cite{bernauer10}, the correction calculated for a
structureless proton, corresponding to the $Q^2 = 0$ limit of the 2nd Born
calculation~\cite{arrington11b}.  Note that this means that no $Q^2$
dependence is included for the exchange of the second photon, while the radius
is determined entirely from the $Q^2$ dependence of the form factors.  This
also implies that the correction is of the wrong sign for $Q^2 \gtorder
0.3$~GeV$^2$, where the correction changes sign for both the soft-photon
approximations and the full TPE calculations~\cite{arrington11b}.

Focusing on the low $Q^2$ regime, relevant to the extraction of $R_p$,
Figure~\ref{fig:tpe_compare} shows the range of TPE corrections to the cross
section based on a set of calculations which include both hard and soft TPE
and which aim to be reliable at low $Q^2$.  These include hadronic calculations
with only protons in the intermediate state by Blunden, Melnitchouk, and
Tjon~\cite{blunden05a} and by  Borisyuk and Kobushkin~\cite{borisyuk06a}, as
well as calculations including intermediate $\Delta$~\cite{kondratyuk05} and
$\pi$-N intermediate states~\cite{borisyuk13}.  Also shown are the Borisyuk
and Kobushkin low-$Q^2$ TPE expansion~\cite{borisyuk07a} and their dispersion
calculations excluding~\cite{borisyuk08} and including~\cite{borisyuk12}
$\Delta$ resonance contributions. While the correction decreases with
increasing $Q^2$ for $\varepsilon > 0.4$, the TPE correction for $\varepsilon
< 0.3$ first increases and then decreases. These corrections have been shown
to have an important effect on the form factor extraction at low $Q^2$
values~\cite{arrington07c}, as well as impacting the extracted charge and
magnetization radii of the proton~\cite{blunden05b, zhan11, borisyuk06a,
arrington11c, bernauer11}. Given that present extractions from electron
scattering measurements yield some inconsistency in the magnetic
radius~\cite{ron11, zhan11}, a detailed examination of all corrections at
small $Q^2$ values is important in determining if there are any issues which
could indicate issues with the extracted charge radii.

While calculations of TPE cross section corrections at higher
$Q^2$ values tend to have greater range of results~\cite{carlson07,
arrington11b}, and for polarization observables yield an even larger range of
predictions~\cite{meziane11}, the various approaches which are expected to
be valid at low $Q^2$ indicate minimal model dependence.  The results show a
typical spread between calculations below 0.05\%, suggesting that the TPE
corrections are well known in this region. At these low $Q^2$ values,
deviations from a linear correction in $\varepsilon$ are small except for
extreme $\varepsilon$ values, making tests of the linearity of the Rosenbluth
separation~\cite{tvaskis06, chen07} fairly insensitive to the low $Q^2$
contributions of these calculations.  Note, however, that these TPE
contributions are taken relative to the Mo \& Tsai~\cite{mo69} prescription
for radiative corrections.  The IR-divergent contribution of the TPE
contributions is included in all radiative correction procedures, to cancel
the divergent term in the interference between electron and proton
Bremsstrahlung.  However, different prescriptions can have different finite
residual contributions from TPE, which differ from that of Mo \& Tsai. Almost
all experiments have used the Mo \& Tsai treatment of the TPE contributions,
except for the recent Mainz measurement~\cite{bernauer10, bernauer13}, which
applied the prescription of Maximon and Tjon~\cite{maximon00}.  Therefore, the
TPE contributions in Fig~\ref{fig:tpe_compare} would have to be modified by
the difference between the Mo \& Tsai and the Maximon and Tjon TPE
prescriptions, as shown in Figure 12 of Ref.~\cite{arrington11b}.  However, at
low $Q^2$ values this difference becomes very small and only weakly-dependent
on $Q^2$, and so has little effect on the form factors in the region of
interest for the radius extraction.

Extraction of the radius requires cross sections measured at low $Q^2$,
meaning low electron energies, especially in the case of large-angle
measurements needed to constrain $G_M$ and the magnetization radius. At low
energies, the change in the electron energy due to its interaction with the
Coulomb potential of the nucleus may yield an important change in the
kinematics at the scattering vertex.  In quasi-elastic scattering this effect
is sometimes estimated in the Effective Momentum Approximation
(EMA)~\cite{aste05}, where the acceleration due to the Coulomb field of a
high-Z nucleus can have a significant impact on the e--p scattering
kinematics by shifting the kinematics at the scattering vertex.  This approach
has the advantage of including the Coulomb interaction to all orders in a
semi-classical picture, and can easily be applied to both high- and low-energy
kinematics.  In addition, the shift accounts for the fact that the hard
photon can probe the proton at a different $Q^2$ value, which is not
included in the evaluation of two-photon exchange in calculations which
work in the limit of $q \to 0$ for the second photon.  More details of the
EMA and comparisons to more complete approaches, where available, are given
in Ref.~\cite{aste07, aste08}.

Figure~\ref{fig:cc_2ndborn} shows the fractional correction to the cross
section in the 2nd Born approximation. One can make a crude estimate of the
impact of these corrections on the radius. The RMS radius, $R_p$, is defined
in terms of the low $Q^2$ expansion of the form factor, $G_E(Q^2) \approx 1 -
Q^2 R_p^2 / 6$, with a similar expression for the magnetic
radius.  A proton RMS radius of 0.85~fm yields $G_E(Q^2) \approx 1 -
3 Q^2$, with $Q^2$ in GeV$^2$, or a fractional slope at $Q^2=0$ of roughly
300\%/GeV$^2$.  The charge form factor contribution to the cross section goes
like $G_E^2(Q^2)$, for a slope of roughly 600\%/GeV$^2$ in
the cross section. Coulomb corrections in the 2nd Born approximation change
the cross section by about 0.2\% between $Q^2$=0.01 and 0.03~GeV$^2$
(Fig.~\ref{fig:cc_2ndborn}), yielding a change in the slope of
0.2\%/0.02~GeV$^2$=10\%/GeV$^2$.  This is roughly 2\% of the total slope
introduced by the protons size, corresponding to a 1\% change in the
extracted radius, in good agreement with the observation of a roughly 0.01~fm
change observed when the 2nd Born calculation is applied to the
data~\cite{rosenfelder00}.  Note that using this measure, the discrepancy
between electron- and muon-based results yields a change in the slope of the
reduced cross section of $\approx$50\%/GeV$^2$, or a 1\% change in the cross
section at $Q^2 = 0.02$~GeV$^2$.

In this paper, we estimate the impact of higher-order Coulomb corrections by
applying the EMA prescription of Ref.~\cite{arrington04c} to elastic e--p
scattering at low $Q^2$. The key parameter in the calculation is the
Coulomb potential at the point where the scattering occurs.  When the EMA is
used to evaluate scattering from a heavy nucleus, it is generally assumed that
the scattering occurs uniformly within the nucleus, so the Coulomb
potential is taken to be something between the surface and central potential.
Averaging over the nuclear volume, assuming a uniform charge sphere, yields a
potential that is 80\% of the central potential, in good agreement with the
result of Ref.~\cite{aste05} which adjusts the average potential in the EMA
calculations to match distorted-wave Born approximation results~\cite{kim96,
udias93} which are more reliable but can only be performed at lower $Q^2$
values.

\begin{figure}[!htbp]
\begin{center}
\includegraphics*[width=3.1in]{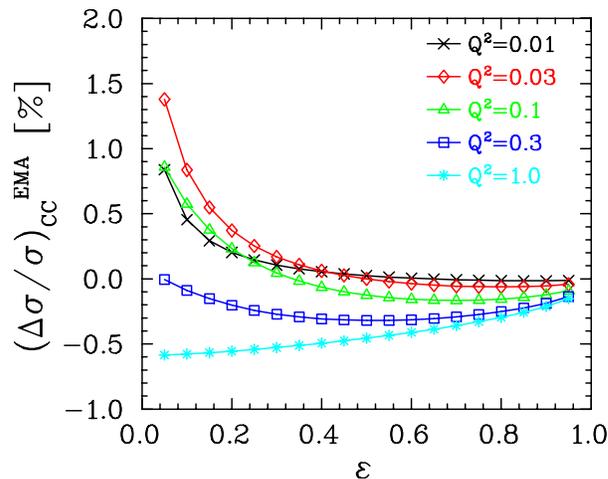}
\end{center}
\caption{Coulomb corrections calculated in the Effective
Momentum Approximation (as described in the text) for e--p elastic scattering
for $0.01 < Q^2 < 1$~GeV$^2$.}
\label{fig:cc_ema}
\end{figure}

At high energies, the EMA calculation yields a correction that is very similar
to the 2nd Born approximation for $\varepsilon > 0.5$, as shown in
Ref.~\cite{arrington04c}, with small differences larger-angle scattering.
Note that the EMA calculation applied in Ref.~\cite{arrington04c} used the
central potential and did not include the flux factor of Ref.~\cite{aste05}.
Inclusion of this correction yields somewhat improved agreement at low
$\varepsilon$, but there are still clear differences remaining.  For heavy
nuclei and low energy scattering, there can be significant differences between
the EMA result and more complete calculations~\cite{aste08}, although the
qualitative behavior is always reasonable.

For e--p elastic scattering at very low $Q^2$, one expects that the scattering
will occur when the electron is outside of the proton, so using the average
Coulomb potential in the proton would be an overestimate. We take the Coulomb
potential corresponding to the case where the scattering occurs at a
separation $\Delta x = 1 / q$, where $q$ is the momentum of the exchanged
virtual photon.  This will suppress the impact of the energy shift for $Q^2$
values where the scattering occurs outside of the proton. This occurs for
$Q^2 < 0.03$~GeV$^2$ if we take the proton to be a uniform sphere of radius
1.15~fm to match the observed RMS radius.  For $Q^2$ values where the
scattering occurs inside of the proton, we limit the Coulomb potential to a
maximum of the volume-averaged value of 1.5 MeV.

Figure~\ref{fig:cc_ema} shows the Coulomb correction in the EMA. The behavior
is qualitatively similar to the 2nd Born approximation result for
$Q^2=1$~GeV$^2$, as it is for larger $Q^2$ values~\cite{arrington04c}. The EMA
result has a very different angular dependence at very low $Q^2$ values, with
a sharp rise at low $\varepsilon$.  As such, this correction could have an
important impact on the low $Q^2$ extraction of the form factors and radius,
especially for the extraction of $G_M$ and the magnetic radius. Note that the
comparison of our EMA and the 2nd Born approximation
(Figs.~\ref{fig:cc_2ndborn} and~\ref{fig:cc_ema}) show a much smaller
correction in the EMA approach at very low $Q^2$ and large $\varepsilon$
(corresponding to larger electron energies), where the EMA should still be
reasonably reliable.  This suggests that taking $\Delta x = 1 / q$ yields too
much suppression of the correction at very low $Q^2$, and that the effects may
be larger in this region.  While the EMA is not expected to yield precise
results in this region, it does allow for contributions that are not included
in the 2nd Born approximation, and is therefore a reasonable tool for an
initial investigation of such effects.

\begin{figure}[!htbp]
\begin{center}
\includegraphics*[width=3.1in]{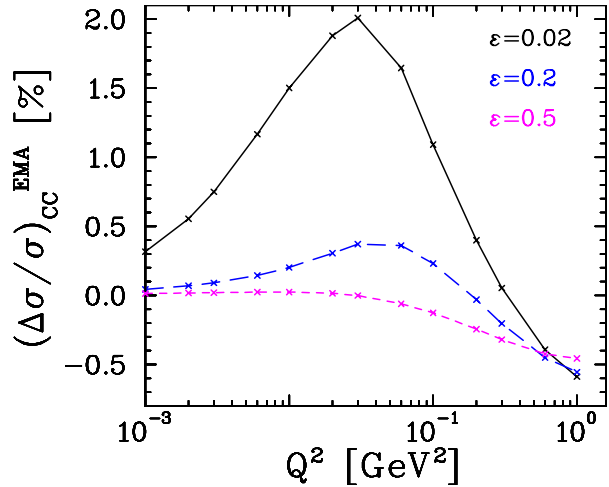}
\end{center}
\caption{$Q^2$ dependence of the EMA correction at
$\varepsilon$=0.02 (solid), 0.2 (long dash), and 0.5 (short dash).}
\label{fig:qsqdep}
\end{figure}

Figure~\ref{fig:qsqdep} shows the EMA result as a function of $Q^2$ for three
$\varepsilon$ values.  At larger $\varepsilon$ values, the effect is small,
especially near $Q^2$=0.  As noted above, explaining the charge radius
discrepancy would require a change in the slope of the reduced cross
section at large $\varepsilon$ of 50\%/GeV$^2$, or a 1\% change between 
$Q^2 = 0.01$ and 0.03~GeV$^2$, while the calculation yields a change of
around 0.05\%.  So this suggests a negligible effect on the extraction
of the charge radius.

For the magnetic radius, the low-$\varepsilon$ results are more significant.
At $\varepsilon=0.02$, there is a very rapid rise with $Q^2$ (note that $Q^2$
is shown on a logarithmic scale). From the $Q^2$ dependence of the correction,
it is straightforward to estimate what impact this would have on a direct
extraction of the magnetic radius from very low $Q^2$ data at $\varepsilon
\approx 0$. The EMA yields a 1\% change in the correction for a change in
$Q^2$ of less than 0.01~GeV$^2$, or a change in the slope of 100\%/$Q^2$. This
would correspond to a very large correction to the magnetic radius.

However, existing measurements below $Q^2$=0.1~GeV$^2$ are limited and there
aren't measurements over this $Q^2$ and $\varepsilon$ range.  The kinematics
of interest for existing data are at somewhat larger $Q^2$ or $\varepsilon$
values. At $\varepsilon$=0.02, the correction decreases by 1.5\% going from
$Q^2=0.03$~GeV$^2$ to 0.2~GeV$^2$, while for $\varepsilon=0.2$, it increases
0.3\% in going from 0.01 to 0.03~GeV$^2$. These provide changes in the
fractional slope of the cross section of -8\%/GeV$^2$ and +15\%/GeV$^2$,
respectively, corresponding to small but potentially significant corrections
to the extracted magnetic proton radius.  In addition, because the extraction
of $G_M$ comes from an extrapolation of the reduced cross section to
$\varepsilon=0$, where the cross section is very small, the fact that the
different $Q^2$ dependence seen at low and high values of $\varepsilon$ may
yield a magnification of the impact on the magnetic radius.

While this rough estimate shows that these effects could have important
contributions to the extraction of the magnetic radius, it is difficult to
provide a more detailed estimate of the impact.  First, because it is
very sensitive to the exact kinematics of the data included in the extraction.
The corrections would have to be applied to each measurement and then the
radius extraction procedure, including fitting of normalization factors if
they are allowed to vary, would have to be repeated.  In addition, it is not
clear how one would combine the results of the EMA at very low $Q^2$ with the
TPE calculations, to include the full impact of the corrections without double
counting.  A more detailed calculation, e.g. in the distorted-wave Born
approximation, would allow for a more reliable evaluation of the importance
of the higher order contributions.

Two-photon exchange effects have been examined in the comparison of e$^-$--p
and e$^+$--p scattering, but existing data are limited in precision and are
at $Q^2$ values that are too large to have a significant impact on the
radius extraction~\cite{arrington04b}.  New e$^\pm$--p measurements have
made to examine two-photon exchange at Novosibirsk~\cite{VEPP-3, nikolenko10,
gramolin12}, Jefferson Lab~\cite{e07005, weinstein09, bennett12, moteabbed13},
and DESY~\cite{olympus, kohl11, milner12}, but will not extend below
$Q^2=0.5$~GeV$^2$ except at very forward angles.  However, these effects can be
examined in future measurements at very low $Q^2$ in comparisons of e$^\pm$--p
and $\mu^\pm$--p) scattering in the MUSE experiment at PSI~\cite{MUSE}.

In conclusion, we find a clear qualitative difference between the 2nd Born
approximation and the EMA evaluation of Coulomb corrections for e--p elastic
scattering at low $Q^2$. The difference is particularly important large
angles, where the data are sensitive to $G_M(Q^2)$, suggesting that effects
beyond two-photon exchange may be important in the extraction of the magnetic
radius.  A quantative estimate of their impact would involve a more detailed
calculation, with the impact of the correction evaluated using on the exact
data set and fitting procedure used to extract the radius.

\ack

I thank M. K. Medina checking the calculations presented here, and A. Aste,
P. Blunden and B. Kobushkin for providing calculations and useful discussion.
This work was supported by the U.S. DOE through contract DE-AC02-06CH11357.

\section*{References}
\bibliographystyle{unsrt_with_abbreviations}
\bibliography{coulomb}

\end{document}